\documentclass[twocolumn,prl,groupedaddress]{revtex4}
\usepackage{graphicx}
\usepackage{epstopdf}
\usepackage{latexsym}
\usepackage{amssymb}
\usepackage{amsfonts} 
\usepackage{txfonts}
\usepackage{pxfonts}
\usepackage[euler]{textgreek}
\usepackage{color}

\graphicspath{{figures/}}

\begin{document}
\title{Reduction in Tension and Stiffening of Lipid Membranes in an Electric Field Revealed by X-ray Scattering}

\author{Arnaud Hemmerle\footnote{Present address: Max Planck Institute for Dynamics and Self-Organization (MPIDS), G\"ottingen 37077, Germany}
}
\affiliation{UPR 22/CNRS, Institut Charles Sadron, Universit\'e de Strasbourg, 23 rue du Loess, BP 84047 67034 Strasbourg Cedex 2, France}

\author{Giovanna Fragneto}
\affiliation{Institut Laue-Langevin, 71 av. des Martyrs, BP 156, 38042 Grenoble Cedex, France}

\author{Jean Daillant}
\affiliation{Synchrotron SOLEIL, L'Orme des Merisiers, Saint-Aubin, BP 48, F-91192 Gif-sur-Yvette Cedex, France}

\author{Thierry Charitat\footnote{thierry.charitat@ics-cnrs.unistra.fr}}
\affiliation{UPR 22/CNRS, Institut Charles Sadron, Universit\'e de Strasbourg, 23 rue du Loess, BP 84047 67034 Strasbourg Cedex 2, France}

\date{\today}

\begin{abstract}

The effect of AC electric fields on the elasticity of supported lipid bilayers has been investigated at the microscopic level using grazing incidence synchrotron x-ray scattering.
A strong decrease in the membrane tension up to 1mN/m and a dramatic increase of its effective rigidity up to 300k$_B$T are observed for local electric potentials seen by the membrane $\lesssim$ 1V.
The experimental results were analyzed using detailed electrokinetic modeling and non-linear Poisson-Boltzmann theory.
Based on a modeling of the electromagnetic stress which provides an accurate description of bilayer separation vs pressure curves, we show that the decrease in tension results from the amplification of charge fluctuations on the membrane surface whereas the increase in bending rigidity results from direct interaction between charges in the electric double layer. These effects eventually lead to a destabilization of the bilayer and vesicle formation.
Similar effects are expected at the tens of nanometer lengthscale in cell membranes with lower tension, and could explain a number of electrically driven processes.
\end{abstract}
\maketitle

\begin{figure}[h]
\centering
\includegraphics[width=.45\textwidth]{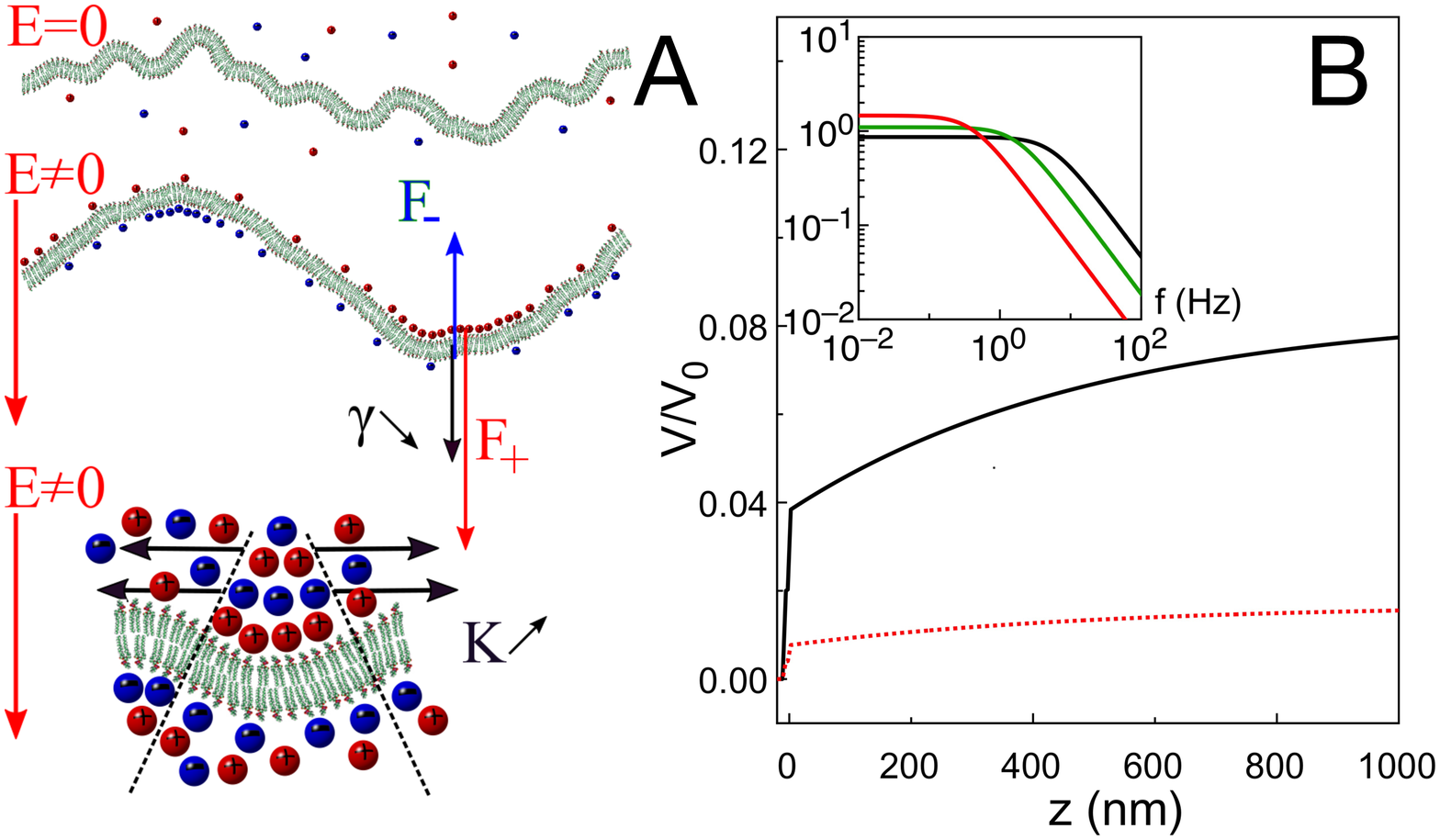}
\caption{Schematic representation of the effect of electric field (A). Top, freely fluctuating bilayer. Middle, interaction between the induced charges and the external electric field leads to an electric force which is amplifying the undulation, acting as a destabilizing negative surface tension. Bottom,  bending brings  the charges closer increasing the electrostatic interactions, mainly in the electric double layer (black arrows) and leads to an increase in bending rigidity. Calculated electric potential as a function of the distance for 10 Hz (black solid curve) and 50 Hz (red dotted line) (B). The inset of (B) shows the local voltage at the membrane boundaries as a function of frequency and for different Debye lengths $\kappa_D^{-1}=800$ nm (red curve), $300$ nm (green curve) and $150$ nm (black curve) using $d_m=5$ nm. $z=0$ corresponds to the middle plane of the floating bilayer.}
\label{fig1}
\end{figure}

Electric fields  can be used to destabilize lipid bilayers as in the electroformation process, the most popular method to form large unilamellar vesicles \cite{Angelova1986}, or to manipulate the shape of vesicles \cite{Riske2005,Dimova2009,Salipante(2015)}. 
Beyond biosensor applications and the investigation of fundamental mechanical, dynamical and binding properties of membranes using impedance spectroscopy or dielectric 
relaxation \cite{Hianik2000}, the strong influence of electric fields on lipid membrane behavior is also used in numerous applications in cell biology, biotechnology and pharmacology \cite{Zhao2006,Zimmermann1986} such as cell hybridization \cite{Zimmermann1996}, electroporation \cite{Son2014}, electrofusion \cite{Rems2013} and electropermeabilization \cite{Vernier2006}. 
All these effects imply a strong deformation of the membranes in the field, the understanding of which in terms of elastic properties is therefore of prime importance \cite{Vlahovska(softmatter2015)}.
Theoretically, the effect of electric fields on membrane tension has been investigated in Ref. \cite{sens2002}, which was extended to bending rigidity in Refs \cite{ziebert2010,ziebert2010b,ziebert2011,Ambjoernsson2007,loubet(PRE2013)}.\\
When placed in an electric field ${\bf E}$, charges of opposite sign will accumulate at both sides of a membrane which can be seen as a capacitor with surface charge densities $\Sigma_\pm$ (see Fig.\ref{fig1}.A), allowing to calculate the normal component of the electromagnetic stress $\left(\Sigma_+^2-\Sigma_-^2\right)/\epsilon_m$ \cite{supplmatt1}. For a flat membrane, a direct consequence is electrostriction: at equilibrium, the elastic response of the membrane (Young modulus $\sim 10^7-10^8$ Pa \cite{Picas2012L01,Hianik2000}) balances the electrostatic pressure \cite{Fadda2013}. Beyond this simple effect, membrane fluctuations modify the boundary conditions for the electric field, leading to a subtle coupling between electrostatics and membrane elasticity. Due to membrane finite thickness $d_m$, a bending deformation induces surface element variations of opposite sign on both interface leading to a net local charge of the bilayer (see Fig.~\ref{fig1}.A, bottom). For a given surface mode $z_q \exp(i {\bf q.r})$, the surface charge density fluctuations are given by $\delta\Sigma_{\pm} = \mp\epsilon_m E_m d_m q^2 z_q\exp(i {\bf q.r})$, where $\epsilon_m$ is the membrane permittivity, $E_m$ the field seen by the membrane, and $q^2 z_q$ the local curvature. Calculating the work of the electromagnetic stress leads to $\delta W\sim-\epsilon_m d_m E_m^2 q^2\left|z_q\right|^2$. As $q^2 \left|z_q\right|^2$ is the increase in area of the fluctuating membrane, this means that there is a negative correction to the free energy, equivalent to a negative (destabilizing) contribution $\Gamma_{m}$ to the membrane surface tension $\gamma$ \cite{sens2002,supplmatt1}.
Similar effects occur in the electric double layer leading to a total correction $\Gamma_{el}=\Gamma_m + \Gamma_{DL}$, where $\Gamma_{DL}$ is the usually smaller correction coming from the electrical double layer \cite{ziebert2010b,ziebert2011,Ambjoernsson2007,loubet(PRE2013),supplmatt1}.
Taking into account non-linear effects in the electric double layer we have,
\begin{equation} 
\Gamma_m = -\epsilon_m d_m E_m^2 =  -\frac{\epsilon_m}{d_m}\left[V_{\rm loc}-\frac{4k_BT}{e}\ln{\left(\frac{1+c}{1-c}\right)}\right]^2,
\label{eq1}
\end{equation}
where $V_{\rm loc}$ is the local electric potential seen by the bilayer and the double electric layer, lower than the applied potential (Fig.\ref{fig1}.B). $0 < c < 1$ is a dimensionless parameter depending on Debye length $\kappa_D^{-1}$ and voltage, saturating to 1 for either high salt concentration or high voltage because of non-linear effects in the double electric layer \cite{ziebert2010b,ziebert2011,supplmatt1}.
Further development in powers of $q$ give contributions in $q^4$ \cite{ziebert2010b,ziebert2011,Ambjoernsson2007,Salipante(2015)}, corresponding to a positive correction $K_{el}=K_m+K_{DL}$ to the membrane bending rigidity $\kappa$: bending brings the charges closer and increases the electrostatic repulsion (see Fig.\ref{fig1}.A, bottom part). Consistently, the largest correction is now due to the thick double layer and is proportional to $\kappa_D^{-1}$:
\begin{equation}
K_{DL} = 4 \epsilon_w \left(\frac{k_B T}{e}\right)^2 \kappa_D^{-1} \frac{c^2(3-c^2)}{1+c^2},
\label{eq2}
\end{equation} 
where $\epsilon_w$ is the permittivity of water.

\begin{figure}[h]
\centering
\includegraphics[width=.45\textwidth]{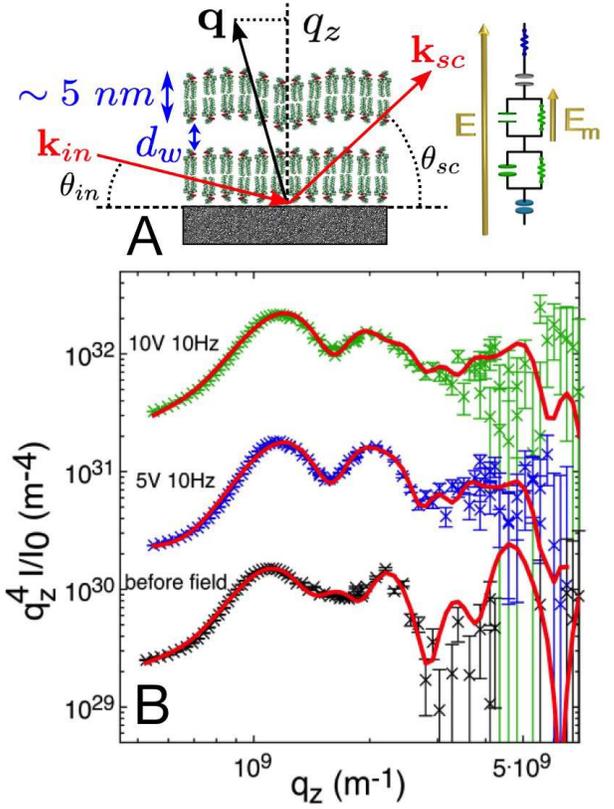}
\caption{Schematic view of the experiment and electrokinetic model of the supported bilayers (A). The incident beam at grazing incidence (direction ${\bf k_{in}}$) is scattered in direction ${\bf k_{sc}}$, giving access to in-plane fluctuations. The off-specular reflectivity curves and the associated best fits (B).}
\label{fig2}
\end{figure}

The model system we have used consisted of two supported bilayers of DSPC ($\scriptstyle{L{-}\alpha}$ 1,2-distearoyl-sn-glycero-3-phosphocholine, Avanti Polar Lipids, Lancaster, Alabama) deposited on ultra-flat silicon substrates (Fig.~\ref{fig2}.A) \cite{charitat1999}. All the experiments were performed in fluid phase at 58$^\circ$C. The first bilayer serves as a spacer to reduce the interaction between the floating second bilayer and the substrate and keeps it free to fluctuate \cite{daillant2005,Hemmerle(PNAS2012)}. Potential was applied between a Cu layer deposited at the back of the thick Si substrate and an ITO coated glass plate mounted parallel to the substrate, 0.5 cm from the membrane.

We used a ${27}$ keV x-ray beam (wavelength ${\lambda=0.0459}$ nm) at the CRG-IF beamline of the European Synchrotron Radiation Facility (ESRF) in an off-specular geometry described in Fig.~\ref{fig2}.A \cite{supplmatt0}. The grazing angle of incidence is kept fixed ($\theta_{in} = 0.7$ mrad), below the critical angle for total external reflection $\theta_c \simeq 0.85$ mrad  as this allows easy background subtraction, and $\theta_{sc}$ is scanned in the plane of incidence \cite{malaquin10, Hemmerle(PNAS2012)}. In all experiments, the incident beam was $500 \muup$m $\times 20 \muup$m and the reflected intensity was defined using a 20 mm$\times 200 \muup$m slit at 210 mm from the sample and a 20 mm$\times 200 \muup$m slit at 815 mm from the sample and recorded using a NaI (Tl) scintillator.

\begin{figure}[h!]
\centering
\includegraphics[width=.49\textwidth]{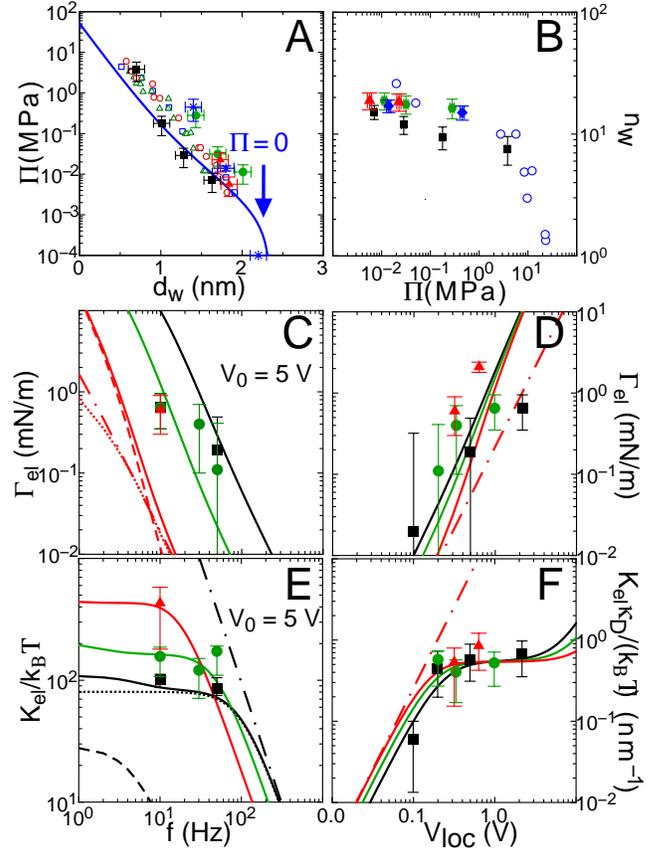}
\caption{Effect of an AC field on a supported bilayer. Filled symbols are data from this work. Black squares (\textcolor{black}{$\blacksquare$}), green circles (\textcolor{green}{$\bullet$}) and red triangles (\textcolor{red}{$\blacktriangle$}) correspond to different experiments. Solid lines correspond to the full electrostatic contribution (Poisson-Boltzmann) to $\Gamma_{el}$ and $K_{el}$ which can be decomposed in a membrane contribution (dotted line) and an electric double layer contribution (dashed line). Linear Debye-H\"uckel theory is shown as dashed-dotted lines. $\kappa_D^{-1}=800$ nm (red curves), $\kappa_D^{-1}=300$ nm (green curves) and $\kappa_D^{-1}=150$ nm (black curves). Mechanical pressure as a function of distance (A). Blue stars are from Ref. \cite{Hemmerle(PNAS2012)} where pressure was applied by osmotic stress on similar double bilayers. Empty symbols are from Ref. \cite{petrache(pre1998)} (osmotic stress on multilayer stacks). The solid line is calculated after Ref. \cite{Hemmerle(PNAS2012)} using dispersive, electrostatic and entropic contributions to the inter-bilayer potential. Number of water molecules per lipid $n_w$ as a function of electrostatic pressure $\Pi$ (B). 
Blue circles (\textcolor{blue}{$\circ$}) were obtained by NMR spectroscopy for osmotically stressed DMPC multilayer stacks\cite{mallikarjunaiah2011solid}. Electrostatic contribution to the membrane tension $\Gamma_{el}$ as a function of frequency for a fixed voltage $V_0=5$ V (C) and as a function of the local voltage $V_{\rm loc}$ at the membrane (D). Electrostatic contribution to the membrane rigidity $K_{el}$ as a function of frequency for a fixed voltage $V_0=5$ V (E) and $K_{el} \kappa_D /k_BT$ as a function of the local potential difference $V_{\rm loc}$ (F).
}
\label{fig3}
\end{figure}

Off-specular scattering is sensitive to both static deformation and thermal fluctuations of bilayers. In the limit of small amplitudes, and the simple case of a single bilayer in  an interaction potential $U$, the scattered intensity is $ I_{sc} \propto\langle z_q z_{-q}\rangle$, with the fluctuation spectrum $\langle z_q z_{-q}\rangle = k_BT/h({\bf q})$ and 
\begin{equation}
h\left({\bf q}\right)=U^{\prime\prime}+\gamma q^2+\kappa q^4.
\label{f(q)}
\end{equation}
The Hamiltonian of the system is given by ${\cal H}=\sum_q {\cal H}({\bf q}) = \sum_q h\left({\bf q}\right)\left|z_q\right|^2$. Fitting of the scattering curves, accounting for both thermal fluctuations and the {\it static} roughness induced by the substrate \cite{andelman2001} (following the procedure described in details in \cite{daillant2005,malaquin10}), gives access to the bilayer electron density profile, $\gamma$, $\kappa$ and $U^{\prime\prime}$.
Different scattering curves are presented in Figure \ref{fig2}.B and in Supplemental Material, showing that high voltage differences, up to 10 V can be applied to the cell without destroying the membrane, but strongly affecting its fluctuations. Fig.~\ref{fig3} summarizes the main findings of this paper.
We clearly observe an electrostriction effect on structural properties. The thickness $d_w$ of the water layer in between the two lipid bilayers decreases with the electric field (Fig.~\ref{fig3}.A). Depending on voltage and frequency, we also observe large negative corrections to the tension $\Gamma_{el} = \gamma_{(V=0)} -\gamma$ (up to 1 mN/m, Fig.~\ref{fig3}.C,D) and positive corrections to the bending stiffness $K_{el} = \kappa -\kappa_{(V=0)}$ (up to a few hundreds of $k_BT$, Fig.~\ref{fig3}.E,F). The measured values $\gamma_{(V=0)}=0.5-1$ mN/m and $\kappa_{(V=0)}=15-20k_B T$ are in good agreement with known values for DSPC bilayer \cite{malaquin10}.

Analyzing our results first requires to determine the local voltage drop $V_{\rm loc}$ seen by the bilayer. To this end, we model the system electrokinetics by solving the Poisson-Nernst-Planck equations, generalizing the model of Ziebert et al \cite{ziebert2011} to the double supported bilayer (Fig.~\ref{fig2}.A) \cite{supplmatt1}. The only unknown parameter is the Debye screening length $\kappa_D^{-1}$, which might slightly depend on the dissolved carbon dioxide and fixes the conductivity of the solution \cite{Haughey1998217}. Whereas $\kappa_D^{-1}$ = 960 nm in pure water, it is reduced to 150 nm for normal atmospheric conditions. As scattering curves were recorded 5 to 10 hours after sample preparation, which can influence the Debye length, $\kappa_D^{-1}$ = 150 nm, 300 nm and 800 nm were used in the analysis. With these values and a single diffusion coefficient $D = 7.5 \times 10^{-9}$ m$^2$/s for all ions \cite{haynes2013crc}, the effective membrane resistance, lower than its intrinsic resistance, ranges from 20 $\Omega.$cm$^2$ to 300 $\Omega.$cm$^2$. The system behaves as a low-pass filter with a cut-off frequency determined by the bulk solution conductance $R_B^{-1}$ and the electric double layer capacitance per unit area $C_{DL}$ (inset of Fig.~\ref{fig1}.B), the highest resistance and capacitance in the system respectively. Depending on Debye length, $R_B = 0.5 - 10$ M$\Omega.$cm$^2$ and $C_{DL} =0.04 - 0.18$ \textmu F/cm$^2$, leading to cutoff frequencies of 0.2 Hz for $\kappa_D^{-1}$ = 150 nm to 3 Hz for $\kappa_D^{-1}$ = 800 nm. Accordingly, the voltage drop at the membrane increases from less than $0.01 V_0$ at 50 Hz to $\approx 0.04 V_0$ at 10 Hz, where $V_0$ is the AC field applied to the membrane (Fig.~\ref{fig2}.A).

First discussing electrostriction, the most compressible part in the system is the water layer in between the two lipid bilayers and the electromagnetic stress is balanced by the interbilayer potential. By plotting the electrostatic pressure $\Pi$ \cite{supplmatt1} as a function of the interbilayer water thickness $d_w$ (Fig.~\ref{fig3}.A), all points fall on a master curve obtained for both the natural entropic repulsion between bilayers \cite{Hemmerle(PNAS2012)} and osmotic stress, either applied on floating bilayers \cite{Hemmerle(PNAS2012)} or multilayer stacks \cite{petrache(pre1998)}, demonstrating that the local electromagnetic stress is well described by our model. We also report in Fig.~\ref{fig3}.B the number of water molecules per lipids $n_w$ \cite{NagleWiener89} as a function of the pressure $\Pi$. Similar curves obtained when the pressure is osmotically applied on a floating bilayer \cite{Hemmerle(PNAS2012)} and on multilayer stacks \cite{mallikarjunaiah2011solid} are also presented, clearly demonstrating that the floating bilayers behave the same way irrespective of how the mechanical stress is applied and keep their integrity under the applied electric field.\\

\begin{figure}[h]
\centering
\includegraphics[width=.49\textwidth]{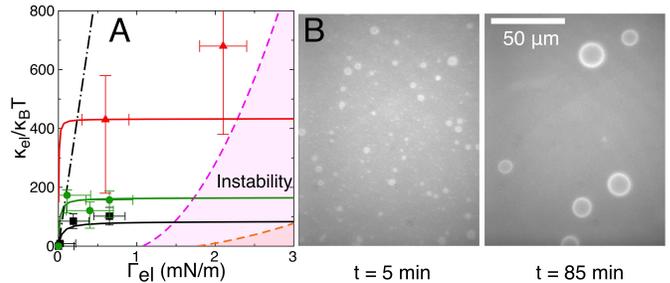}
\caption{$\Gamma_{el}$ as a function of $K_{el}$ (A). The destabilization limits for $U^{\prime\prime}=10^{12.5}$ J.m$^{-4}$ and $10^{11.5}$ J.m$^{-4}$ are given as light and dark pink domains respectively. Vesicle formation under electric field (5 V, 5 Hz) from a single supported bilayer of DPPC (B). Observation by fluorescence microscopy at 5 min (left) and 85 min (right) after the application of the field.}
\label{fig4}
\end{figure}

The frequency dependence of the correction to the membrane tension $\Gamma_{el}$ is plotted in Fig.~\ref{fig3}.C for $V_0=5$ V, where a  $\approx \omega^{-2}$ decay is observed. The origin of this purely electrokinetic effect lies in the impossibility to charge the membrane above the cutoff frequency of the low bandpass filter formed by the electric double layer capacitor and bulk water resistor due to the finite mobility of ions in water.
By plotting $\Gamma_{el}$ as a function of the local electric field $V_{\rm loc}$, we observe a good agreement between data and theoretical predictions with $\Gamma_{el}$ exhibiting a roughly $\propto V^2_{\rm loc}$ dependence (Fig. \ref{fig3}D).\\
The increase in bending rigidity $K_{el}$ is plotted as a function of frequency for $V_0=5$ V in Fig.~\ref{fig3}.E and as a function of $V_{\rm loc}$ in Fig.~\ref{fig3}.F. Both curves exhibit a more complex behavior than the $\Gamma_{el}$ curves which can be attributed to non-linear effects due to the large voltage drop at the membrane with $eV/k_B T \simeq 1$ (Fig.~\ref{fig1}.B).
In contrast with the linear theory which exhibits the expected $\omega^{-2}$ behavior, the low-frequency plateau seen for both experimental data and non-linear theory in Fig.~\ref{fig3}.E comes from saturation effects in the electric double layer.
By plotting the data as a function of $V_{\rm loc}$, which allows one to decouple microscopic and electrokinetic effects, all $K_{el}\kappa_D$ values indeed fall on a master curve with a saturation from 0.5V (see Fig. \ref{fig3}.F). This is in remarkable agreement with the theory which predicts a saturation value of $K_{el}$ proportional to the Debye length \cite{ziebert2011}, and fully consistent with the expectation that a thicker layer is more difficult to bend.
As $\kappa_D$ also fixes independently cutoff frequencies via water conductivity, the analysis is clearly consistent.
However, we must point out that despite its remarkable description of our data, the theory of Ref. \cite{ziebert2010b} is for a single bilayer in a symmetric environment, unlike the experimental conditions used here.\\
The electroformation technique uses similar electric field to destabilize membranes and fabricate Giant Unilamellar Vesicles (GUVs).
The stability limit of the bilayers can be calculated using $h({\bf q}) = 0$,
and is drawn in Fig.~\ref{fig4}.A for two different values of the potential second derivative ($U^{\prime\prime}=3 \times 10^{11}$ J.m$^{-4}$ and $3 \times 10^{12}$ J.m$^{-4}$). It clearly shows that our x-rays experiments are performed in the stability domain but close to instability conditions. With the aim of observing destabilization, we have applied an electric field on a single supported bilayer on an ITO coated glass slide under similar conditions. 
We observed by fluorescence microscopy the formation of GUVs above and close to the main transition temperature $T_m$ (Fig.~\ref{fig4}.B). Small vesicles of diameter $\approx$5 \textmu m are the dominant population at short times ($t \sim$ 1-10 min) and grow with time to reach a diameter of 10-30 \textmu m.  Interestingly, the initial size we find here is consistent with the instability in the bilayer fluctuation spectra evidenced by x-ray scattering.

The effect of AC fields on supported floating bilayers has been investigated by x-ray off-specular scattering. In a consistent set of experimental data strongly supported by an established theoretical model, we have evidenced both a dramatic decrease in the membrane tension, possibly down to negative values, and a strong increase in the bilayer bending rigidity. We demonstrate that the effect on tension results from an amplification of charge fluctuations at the membrane.The effect on rigidity comes from couplings inside the electric double layer, and can only be understood by using the full non- linear Poisson-Boltzmann theory. The effect of voltage and AC field frequency has been characterized. The competition between the stabilizing effect on bending rigidity (mainly acting at lengthscales $\leq 0.5 \mu m$) and destabilizing effect on tension (at lengthscales $\geq 0.5 \mu m$) leads to $\approx 1 \mu m$ vesicle formation as observed. This detailed understanding can now be used for further analysis of the effect of electric fields on biological membranes. For cell membranes which have a smaller rigidity ($\sim 1-10$ k$_B$T) than our model membrane, destabilization is expected to occur at lengthscales $\simeq 50$ nm and could explain the effect of low electric fields in processes like electroendocytosis.

\vspace{-0.5cm} 
\section*{Acknowledgments}
The authors wish to thank L. Malaquin and S. Micha for assistance during the experiments and L. Malaquin, D. Lacoste and F. Ziebert for useful discussions. Supports from the Labex NIE 11-LABX-0058-NIE (Investissement d'Avenir programme ANR-10- IDEX-0002-02) and PSCM facilities at the ILL for sample preparation are gratefully acknowledged.


\begin{thebibliography}{32}%
\makeatletter
\providecommand \@ifxundefined [1]{%
 \@ifx{#1\undefined}
}%
\providecommand \@ifnum [1]{%
 \ifnum #1\expandafter \@firstoftwo
 \else \expandafter \@secondoftwo
 \fi
}%
\providecommand \@ifx [1]{%
 \ifx #1\expandafter \@firstoftwo
 \else \expandafter \@secondoftwo
 \fi
}%
\providecommand \natexlab [1]{#1}%
\providecommand \enquote  [1]{``#1''}%
\providecommand \bibnamefont  [1]{#1}%
\providecommand \bibfnamefont [1]{#1}%
\providecommand \citenamefont [1]{#1}%
\providecommand \href@noop [0]{\@secondoftwo}%
\providecommand \href [0]{\begingroup \@sanitize@url \@href}%
\providecommand \@href[1]{\@@startlink{#1}\@@href}%
\providecommand \@@href[1]{\endgroup#1\@@endlink}%
\providecommand \@sanitize@url [0]{\catcode `\\12\catcode `\$12\catcode
  `\&12\catcode `\#12\catcode `\^12\catcode `\_12\catcode `\%12\relax}%
\providecommand \@@startlink[1]{}%
\providecommand \@@endlink[0]{}%
\providecommand \url  [0]{\begingroup\@sanitize@url \@url }%
\providecommand \@url [1]{\endgroup\@href {#1}{\urlprefix }}%
\providecommand \urlprefix  [0]{URL }%
\providecommand \Eprint [0]{\href }%
\providecommand \doibase [0]{http://dx.doi.org/}%
\providecommand \selectlanguage [0]{\@gobble}%
\providecommand \bibinfo  [0]{\@secondoftwo}%
\providecommand \bibfield  [0]{\@secondoftwo}%
\providecommand \translation [1]{[#1]}%
\providecommand \BibitemOpen [0]{}%
\providecommand \bibitemStop [0]{}%
\providecommand \bibitemNoStop [0]{.\EOS\space}%
\providecommand \EOS [0]{\spacefactor3000\relax}%
\providecommand \BibitemShut  [1]{\csname bibitem#1\endcsname}%
\let\auto@bib@innerbib\@empty
\bibitem [{\citenamefont {Angelova}\ and\ \citenamefont
  {Dimitrov}(1986)}]{Angelova1986}%
  \BibitemOpen
  \bibfield  {author} {\bibinfo {author} {\bibfnamefont {M.}~\bibnamefont
  {Angelova}}\ and\ \bibinfo {author} {\bibfnamefont {D.}~\bibnamefont
  {Dimitrov}},\ }\href@noop {} {\bibfield  {journal} {\bibinfo  {journal}
  {Faraday Discuss. Chem. Soc.}\ }\textbf {\bibinfo {volume} {81}},\ \bibinfo
  {pages} {303} (\bibinfo {year} {1986})}\BibitemShut {NoStop}%
\bibitem [{\citenamefont {Riske}\ and\ \citenamefont
  {Dimova}(2005)}]{Riske2005}%
  \BibitemOpen
  \bibfield  {author} {\bibinfo {author} {\bibfnamefont {K.~A.}\ \bibnamefont
  {Riske}}\ and\ \bibinfo {author} {\bibfnamefont {R.}~\bibnamefont {Dimova}},\
  }\href@noop {} {\bibfield  {journal} {\bibinfo  {journal} {Biophysical
  Journal}\ }\textbf {\bibinfo {volume} {88}},\ \bibinfo {pages} {1143 }
  (\bibinfo {year} {2005})}\BibitemShut {NoStop}%
\bibitem [{\citenamefont {Dimova}\ \emph {et~al.}(2009)\citenamefont {Dimova},
  \citenamefont {Bezlyepkina}, \citenamefont {Domange~Jordo}, \citenamefont
  {Knorr}, \citenamefont {Riske}, \citenamefont {Staykova}, \citenamefont
  {Vlahovska}, \citenamefont {Yamamoto}, \citenamefont {Yang},\ and\
  \citenamefont {Lipowsky}}]{Dimova2009}%
  \BibitemOpen
  \bibfield  {author} {\bibinfo {author} {\bibfnamefont {R.}~\bibnamefont
  {Dimova}}, \bibinfo {author} {\bibfnamefont {N.}~\bibnamefont {Bezlyepkina}},
  \bibinfo {author} {\bibfnamefont {M.}~\bibnamefont {Domange~Jordo}}, \bibinfo
  {author} {\bibfnamefont {R.}~\bibnamefont {Knorr}}, \bibinfo {author}
  {\bibfnamefont {K.}~\bibnamefont {Riske}}, \bibinfo {author} {\bibfnamefont
  {M.}~\bibnamefont {Staykova}}, \bibinfo {author} {\bibfnamefont
  {P.}~\bibnamefont {Vlahovska}}, \bibinfo {author} {\bibfnamefont
  {T.}~\bibnamefont {Yamamoto}}, \bibinfo {author} {\bibfnamefont
  {P.}~\bibnamefont {Yang}}, \ and\ \bibinfo {author} {\bibfnamefont
  {R.}~\bibnamefont {Lipowsky}},\ }\href@noop {} {\bibfield  {journal}
  {\bibinfo  {journal} {Soft Matter}\ }\textbf {\bibinfo {volume} {5}},\
  \bibinfo {pages} {3201} (\bibinfo {year} {2009})}\BibitemShut {NoStop}%
\bibitem [{\citenamefont {Salipante}\ \emph {et~al.}(2015)\citenamefont
  {Salipante}, \citenamefont {Shapiro},\ and\ \citenamefont
  {Vlahovska}}]{Salipante(2015)}%
  \BibitemOpen
  \bibfield  {author} {\bibinfo {author} {\bibfnamefont {P.~F.}\ \bibnamefont
  {Salipante}}, \bibinfo {author} {\bibfnamefont {M.~L.}\ \bibnamefont
  {Shapiro}}, \ and\ \bibinfo {author} {\bibfnamefont {P.~M.}\ \bibnamefont
  {Vlahovska}},\ }\href@noop {} {\bibfield  {journal} {\bibinfo  {journal}
  {Procedia IUTAM}\ }\textbf {\bibinfo {volume} {16}},\ \bibinfo {pages} {60 }
  (\bibinfo {year} {2015})}\BibitemShut {NoStop}%
\bibitem [{\citenamefont {Hianik}(2000)}]{Hianik2000}%
  \BibitemOpen
  \bibfield  {author} {\bibinfo {author} {\bibfnamefont {T.}~\bibnamefont
  {Hianik}},\ }\href@noop {} {\bibfield  {journal} {\bibinfo  {journal}
  {Reviews in Molecular Biotechnology}\ }\textbf {\bibinfo {volume} {74}},\
  \bibinfo {pages} {189 } (\bibinfo {year} {2000})}\BibitemShut {NoStop}%
\bibitem [{\citenamefont {Zhao}\ \emph {et~al.}(2006)\citenamefont {Zhao},
  \citenamefont {Song}, \citenamefont {Pu}, \citenamefont {Wada}, \citenamefont
  {Reid}, \citenamefont {Tai}, \citenamefont {Wang}, \citenamefont {Guo},
  \citenamefont {Walczysko}, \citenamefont {Gu}, \citenamefont {Sasaki},
  \citenamefont {Suzuki}, \citenamefont {Forrester}, \citenamefont {Bourne},
  \citenamefont {Devreotes},\ and\ \citenamefont {Penninger}}]{Zhao2006}%
  \BibitemOpen
  \bibfield  {author} {\bibinfo {author} {\bibfnamefont {M.}~\bibnamefont
  {Zhao}}, \bibinfo {author} {\bibfnamefont {B.}~\bibnamefont {Song}}, \bibinfo
  {author} {\bibfnamefont {J.}~\bibnamefont {Pu}}, \bibinfo {author}
  {\bibfnamefont {T.}~\bibnamefont {Wada}}, \bibinfo {author} {\bibfnamefont
  {B.}~\bibnamefont {Reid}}, \bibinfo {author} {\bibfnamefont {G.}~\bibnamefont
  {Tai}}, \bibinfo {author} {\bibfnamefont {F.}~\bibnamefont {Wang}}, \bibinfo
  {author} {\bibfnamefont {A.}~\bibnamefont {Guo}}, \bibinfo {author}
  {\bibfnamefont {P.}~\bibnamefont {Walczysko}}, \bibinfo {author}
  {\bibfnamefont {Y.}~\bibnamefont {Gu}}, \bibinfo {author} {\bibfnamefont
  {T.}~\bibnamefont {Sasaki}}, \bibinfo {author} {\bibfnamefont
  {A.}~\bibnamefont {Suzuki}}, \bibinfo {author} {\bibfnamefont
  {J.}~\bibnamefont {Forrester}}, \bibinfo {author} {\bibfnamefont
  {H.}~\bibnamefont {Bourne}}, \bibinfo {author} {\bibfnamefont
  {C.}~\bibnamefont {Devreotes}, \bibfnamefont {P.N.~McCaig}}, \ and\ \bibinfo
  {author} {\bibfnamefont {J.}~\bibnamefont {Penninger}},\ }\href@noop {}
  {\bibfield  {journal} {\bibinfo  {journal} {Nature}\ }\textbf {\bibinfo
  {volume} {442}},\ \bibinfo {pages} {457} (\bibinfo {year}
  {2006})}\BibitemShut {NoStop}%
\bibitem [{\citenamefont {Zimmermann}(1986)}]{Zimmermann1986}%
  \BibitemOpen
  \bibfield  {author} {\bibinfo {author} {\bibfnamefont {U.}~\bibnamefont
  {Zimmermann}},\ }in\ \href@noop {} {\emph {\bibinfo {booktitle} {Reviews of
  Physiology}}},\ Vol.\ \bibinfo {volume} {105}\ (\bibinfo  {publisher}
  {Springer Berlin Heidelberg},\ \bibinfo {year} {1986})\ pp.\ \bibinfo {pages}
  {175--256}\BibitemShut {NoStop}%
\bibitem [{\citenamefont {Zimmermann}\ and\ \citenamefont
  {Neil}(1996)}]{Zimmermann1996}%
  \BibitemOpen
  \bibfield  {author} {\bibinfo {author} {\bibfnamefont {U.}~\bibnamefont
  {Zimmermann}}\ and\ \bibinfo {author} {\bibfnamefont {G.}~\bibnamefont
  {Neil}},\ }\href@noop {} {\emph {\bibinfo {title} {Electromanipulation of
  cells}}}\ (\bibinfo  {publisher} {CRC press},\ \bibinfo {year}
  {1996})\BibitemShut {NoStop}%
\bibitem [{\citenamefont {Son}\ \emph {et~al.}(2014)\citenamefont {Son},
  \citenamefont {Smith}, \citenamefont {Gowrishankar}, \citenamefont
  {Vernier},\ and\ \citenamefont {Weaver}}]{Son2014}%
  \BibitemOpen
  \bibfield  {author} {\bibinfo {author} {\bibfnamefont {R.~S.}\ \bibnamefont
  {Son}}, \bibinfo {author} {\bibfnamefont {K.~C.}\ \bibnamefont {Smith}},
  \bibinfo {author} {\bibfnamefont {T.~R.}\ \bibnamefont {Gowrishankar}},
  \bibinfo {author} {\bibfnamefont {P.~T.}\ \bibnamefont {Vernier}}, \ and\
  \bibinfo {author} {\bibfnamefont {J.~C.}\ \bibnamefont {Weaver}},\
  }\href@noop {} {\bibfield  {journal} {\bibinfo  {journal} {J Membr Biol}\
  }\textbf {\bibinfo {volume} {247}},\ \bibinfo {pages} {1209} (\bibinfo {year}
  {2014})}\BibitemShut {NoStop}%
\bibitem [{\citenamefont {Rems}\ \emph {et~al.}(2013)\citenamefont {Rems},
  \citenamefont {Usaj}, \citenamefont {Kanduser}, \citenamefont {Rebersek},
  \citenamefont {Miklavcic},\ and\ \citenamefont {Pucihar}}]{Rems2013}%
  \BibitemOpen
  \bibfield  {author} {\bibinfo {author} {\bibfnamefont {L.}~\bibnamefont
  {Rems}}, \bibinfo {author} {\bibfnamefont {M.}~\bibnamefont {Usaj}}, \bibinfo
  {author} {\bibfnamefont {M.}~\bibnamefont {Kanduser}}, \bibinfo {author}
  {\bibfnamefont {M.}~\bibnamefont {Rebersek}}, \bibinfo {author}
  {\bibfnamefont {D.}~\bibnamefont {Miklavcic}}, \ and\ \bibinfo {author}
  {\bibfnamefont {G.}~\bibnamefont {Pucihar}},\ }\href@noop {} {\bibfield
  {journal} {\bibinfo  {journal} {Sci Rep}\ }\textbf {\bibinfo {volume} {3}},\
  \bibinfo {pages} {3382} (\bibinfo {year} {2013})}\BibitemShut {NoStop}%
\bibitem [{\citenamefont {Vernier}\ \emph {et~al.}(2006)\citenamefont
  {Vernier}, \citenamefont {Sun},\ and\ \citenamefont
  {Gundersen}}]{Vernier2006}%
  \BibitemOpen
  \bibfield  {author} {\bibinfo {author} {\bibfnamefont {P.}~\bibnamefont
  {Vernier}}, \bibinfo {author} {\bibfnamefont {Y.}~\bibnamefont {Sun}}, \ and\
  \bibinfo {author} {\bibfnamefont {M.}~\bibnamefont {Gundersen}},\ }\href@noop
  {} {\bibfield  {journal} {\bibinfo  {journal} {BMC Cell Biology}\ }\textbf
  {\bibinfo {volume} {7}},\ \bibinfo {pages} {37} (\bibinfo {year}
  {2006})}\BibitemShut {NoStop}%
\bibitem [{\citenamefont {Vlahovska}(2015)}]{Vlahovska(softmatter2015)}%
  \BibitemOpen
  \bibfield  {author} {\bibinfo {author} {\bibfnamefont {P.~M.}\ \bibnamefont
  {Vlahovska}},\ }\href@noop {} {\bibfield  {journal} {\bibinfo  {journal}
  {Soft Matter}\ }\textbf {\bibinfo {volume} {11}},\ \bibinfo {pages} {7232}
  (\bibinfo {year} {2015})}\BibitemShut {NoStop}%
\bibitem [{\citenamefont {Sens}\ and\ \citenamefont
  {Isambert}(2002)}]{sens2002}%
  \BibitemOpen
  \bibfield  {author} {\bibinfo {author} {\bibfnamefont {P.}~\bibnamefont
  {Sens}}\ and\ \bibinfo {author} {\bibfnamefont {H.}~\bibnamefont
  {Isambert}},\ }\href@noop {} {\bibfield  {journal} {\bibinfo  {journal}
  {Physical Review Letters}\ }\textbf {\bibinfo {volume} {88}},\ \bibinfo
  {pages} {128102} (\bibinfo {year} {2002})}\BibitemShut {NoStop}%
\bibitem [{\citenamefont {Ziebert}\ \emph {et~al.}(2010)\citenamefont
  {Ziebert}, \citenamefont {Bazant},\ and\ \citenamefont
  {Lacoste}}]{ziebert2010}%
  \BibitemOpen
  \bibfield  {author} {\bibinfo {author} {\bibfnamefont {F.}~\bibnamefont
  {Ziebert}}, \bibinfo {author} {\bibfnamefont {M.-Z.}\ \bibnamefont {Bazant}},
  \ and\ \bibinfo {author} {\bibfnamefont {D.}~\bibnamefont {Lacoste}},\
  }\href@noop {} {\bibfield  {journal} {\bibinfo  {journal} {Physical Review
  E}\ }\textbf {\bibinfo {volume} {81}},\ \bibinfo {pages} {031912} (\bibinfo
  {year} {2010})}\BibitemShut {NoStop}%
\bibitem [{\citenamefont {Ziebert}\ and\ \citenamefont
  {Lacoste}(2010)}]{ziebert2010b}%
  \BibitemOpen
  \bibfield  {author} {\bibinfo {author} {\bibfnamefont {F.}~\bibnamefont
  {Ziebert}}\ and\ \bibinfo {author} {\bibfnamefont {D.}~\bibnamefont
  {Lacoste}},\ }\href@noop {} {\bibfield  {journal} {\bibinfo  {journal} {New
  Journal of Physics}\ }\textbf {\bibinfo {volume} {12}},\ \bibinfo {pages}
  {095002} (\bibinfo {year} {2010})}\BibitemShut {NoStop}%
\bibitem [{\citenamefont {Ziebert}\ and\ \citenamefont
  {Lacoste}(2011)}]{ziebert2011}%
  \BibitemOpen
  \bibfield  {author} {\bibinfo {author} {\bibfnamefont {F.}~\bibnamefont
  {Ziebert}}\ and\ \bibinfo {author} {\bibfnamefont {D.}~\bibnamefont
  {Lacoste}},\ }\href@noop {} {\bibfield  {journal} {\bibinfo  {journal}
  {Advances in planar lipid bilayers and liposomes, vol. 14}\ ,\ \bibinfo
  {pages} {63}} (\bibinfo {year} {2011})}\BibitemShut {NoStop}%
\bibitem [{\citenamefont {Ambj\"ornsson}\ \emph {et~al.}(2007)\citenamefont
  {Ambj\"ornsson}, \citenamefont {Lomholt},\ and\ \citenamefont
  {Hansen}}]{Ambjoernsson2007}%
  \BibitemOpen
  \bibfield  {author} {\bibinfo {author} {\bibfnamefont {T.}~\bibnamefont
  {Ambj\"ornsson}}, \bibinfo {author} {\bibfnamefont {M.~A.}\ \bibnamefont
  {Lomholt}}, \ and\ \bibinfo {author} {\bibfnamefont {P.~L.}\ \bibnamefont
  {Hansen}},\ }\href@noop {} {\bibfield  {journal} {\bibinfo  {journal} {Phys.
  Rev. E}\ }\textbf {\bibinfo {volume} {75}},\ \bibinfo {pages} {051916}
  (\bibinfo {year} {2007})}\BibitemShut {NoStop}%
\bibitem [{\citenamefont {Loubet}\ \emph {et~al.}(2013)\citenamefont {Loubet},
  \citenamefont {Hansen},\ and\ \citenamefont {Lomholt}}]{loubet(PRE2013)}%
  \BibitemOpen
  \bibfield  {author} {\bibinfo {author} {\bibfnamefont {B.}~\bibnamefont
  {Loubet}}, \bibinfo {author} {\bibfnamefont {P.~L.}\ \bibnamefont {Hansen}},
  \ and\ \bibinfo {author} {\bibfnamefont {M.~A.}\ \bibnamefont {Lomholt}},\
  }\href@noop {} {\bibfield  {journal} {\bibinfo  {journal} {Phys. Rev. E}\
  }\textbf {\bibinfo {volume} {88}},\ \bibinfo {pages} {062715} (\bibinfo
  {year} {2013})}\BibitemShut {NoStop}%
\bibitem [{sup({\natexlab{a}})}]{supplmatt1}%
  \BibitemOpen
  \href@noop {} {} \bibinfo {note} {see Supplemental
  Material at [] for more details on modeling.}\BibitemShut {Stop}%
\bibitem [{\citenamefont {Picas}\ \emph {et~al.}(2012)\citenamefont {Picas},
  \citenamefont {Rico},\ and\ \citenamefont {Scheuring}}]{Picas2012L01}%
  \BibitemOpen
  \bibfield  {author} {\bibinfo {author} {\bibfnamefont {L.}~\bibnamefont
  {Picas}}, \bibinfo {author} {\bibfnamefont {F.}~\bibnamefont {Rico}}, \ and\
  \bibinfo {author} {\bibfnamefont {S.}~\bibnamefont {Scheuring}},\ }\href@noop
  {} {\bibfield  {journal} {\bibinfo  {journal} {Biophysical Journal}\ }\textbf
  {\bibinfo {volume} {102}},\ \bibinfo {pages} {L01 } (\bibinfo {year}
  {2012})}\BibitemShut {NoStop}%
\bibitem [{\citenamefont {Fadda}\ \emph {et~al.}(2013)\citenamefont {Fadda},
  \citenamefont {Lairez}, \citenamefont {Guennouni},\ and\ \citenamefont
  {Koutsioubas}}]{Fadda2013}%
  \BibitemOpen
  \bibfield  {author} {\bibinfo {author} {\bibfnamefont {G.}~\bibnamefont
  {Fadda}}, \bibinfo {author} {\bibfnamefont {D.}~\bibnamefont {Lairez}},
  \bibinfo {author} {\bibfnamefont {Z.}~\bibnamefont {Guennouni}}, \ and\
  \bibinfo {author} {\bibfnamefont {A.}~\bibnamefont {Koutsioubas}},\
  }\href@noop {} {\bibfield  {journal} {\bibinfo  {journal} {Physical Review
  Letters}\ }\textbf {\bibinfo {volume} {111}},\ \bibinfo {pages} {028102}
  (\bibinfo {year} {2013})}\BibitemShut {NoStop}%
\bibitem [{\citenamefont {Charitat}\ \emph {et~al.}(1999)\citenamefont
  {Charitat}, \citenamefont {Bellet-Amalric}, \citenamefont {Fragneto},\ and\
  \citenamefont {Graner}}]{charitat1999}%
  \BibitemOpen
  \bibfield  {author} {\bibinfo {author} {\bibfnamefont {T.}~\bibnamefont
  {Charitat}}, \bibinfo {author} {\bibfnamefont {E.}~\bibnamefont
  {Bellet-Amalric}}, \bibinfo {author} {\bibfnamefont {G.}~\bibnamefont
  {Fragneto}}, \ and\ \bibinfo {author} {\bibfnamefont {F.}~\bibnamefont
  {Graner}},\ }\href@noop {} {\bibfield  {journal} {\bibinfo  {journal}
  {European Physical Journal B}\ }\textbf {\bibinfo {volume} {8}},\ \bibinfo
  {pages} {583} (\bibinfo {year} {1999})}\BibitemShut {NoStop}%
\bibitem [{\citenamefont {Daillant}\ \emph {et~al.}(2005)\citenamefont
  {Daillant}, \citenamefont {Bellet-Amalric}, \citenamefont {Braslau},
  \citenamefont {Charitat}, \citenamefont {Fragneto}, \citenamefont {Graner},
  \citenamefont {Mora}, \citenamefont {Rieutord},\ and\ \citenamefont
  {Stidder}}]{daillant2005}%
  \BibitemOpen
  \bibfield  {author} {\bibinfo {author} {\bibfnamefont {J.}~\bibnamefont
  {Daillant}}, \bibinfo {author} {\bibfnamefont {E.}~\bibnamefont
  {Bellet-Amalric}}, \bibinfo {author} {\bibfnamefont {A.}~\bibnamefont
  {Braslau}}, \bibinfo {author} {\bibfnamefont {T.}~\bibnamefont {Charitat}},
  \bibinfo {author} {\bibfnamefont {G.}~\bibnamefont {Fragneto}}, \bibinfo
  {author} {\bibfnamefont {F.}~\bibnamefont {Graner}}, \bibinfo {author}
  {\bibfnamefont {S.}~\bibnamefont {Mora}}, \bibinfo {author} {\bibfnamefont
  {F.}~\bibnamefont {Rieutord}}, \ and\ \bibinfo {author} {\bibfnamefont
  {B.}~\bibnamefont {Stidder}},\ }\href@noop {} {\bibfield  {journal} {\bibinfo
   {journal} {The Procedings of the National Academy of Sciences USA}\ }\textbf
  {\bibinfo {volume} {102}},\ \bibinfo {pages} {11639} (\bibinfo {year}
  {2005})}\BibitemShut {NoStop}%
\bibitem [{\citenamefont {Hemmerle}\ \emph {et~al.}(2012)\citenamefont
  {Hemmerle}, \citenamefont {Malaquin}, \citenamefont {Charitat}, \citenamefont
  {Lecuyer}, \citenamefont {Fragneto},\ and\ \citenamefont
  {Daillant}}]{Hemmerle(PNAS2012)}%
  \BibitemOpen
  \bibfield  {author} {\bibinfo {author} {\bibfnamefont {A.}~\bibnamefont
  {Hemmerle}}, \bibinfo {author} {\bibfnamefont {L.}~\bibnamefont {Malaquin}},
  \bibinfo {author} {\bibfnamefont {T.}~\bibnamefont {Charitat}}, \bibinfo
  {author} {\bibfnamefont {S.}~\bibnamefont {Lecuyer}}, \bibinfo {author}
  {\bibfnamefont {G.}~\bibnamefont {Fragneto}}, \ and\ \bibinfo {author}
  {\bibfnamefont {J.}~\bibnamefont {Daillant}},\ }\href@noop {} {\bibfield
  {journal} {\bibinfo  {journal} {The Proceedings of the National Academy of
  Sciences}\ }\textbf {\bibinfo {volume} {109}},\ \bibinfo {pages} {19938}
  (\bibinfo {year} {2012})}\BibitemShut {NoStop}%
\bibitem [{sup({\natexlab{b}})}]{supplmatt0}%
  \BibitemOpen
  \href@noop {} {} \bibinfo {note} {see Supplemental
  Material at [] for more details on experimental methods.}\BibitemShut {Stop}%
\bibitem [{\citenamefont {Malaquin}\ \emph {et~al.}(2010)\citenamefont
  {Malaquin}, \citenamefont {Charitat},\ and\ \citenamefont
  {Daillant}}]{malaquin10}%
  \BibitemOpen
  \bibfield  {author} {\bibinfo {author} {\bibfnamefont {L.}~\bibnamefont
  {Malaquin}}, \bibinfo {author} {\bibfnamefont {T.}~\bibnamefont {Charitat}},
  \ and\ \bibinfo {author} {\bibfnamefont {J.}~\bibnamefont {Daillant}},\
  }\href@noop {} {\bibfield  {journal} {\bibinfo  {journal} {Eur. Phys. J. E}\
  }\textbf {\bibinfo {volume} {31}},\ \bibinfo {pages} {285} (\bibinfo {year}
  {2010})}\BibitemShut {NoStop}%
\bibitem [{\citenamefont {Swain}\ and\ \citenamefont
  {Andelman}(2001)}]{andelman2001}%
  \BibitemOpen
  \bibfield  {author} {\bibinfo {author} {\bibfnamefont {P.~S.}\ \bibnamefont
  {Swain}}\ and\ \bibinfo {author} {\bibfnamefont {D.}~\bibnamefont
  {Andelman}},\ }\href@noop {} {\bibfield  {journal} {\bibinfo  {journal}
  {Physical Review E}\ }\textbf {\bibinfo {volume} {63}},\ \bibinfo {pages}
  {51911} (\bibinfo {year} {2001})}\BibitemShut {NoStop}%
\bibitem [{\citenamefont {Haughey}\ and\ \citenamefont
  {Earnshaw}(1998)}]{Haughey1998217}%
  \BibitemOpen
  \bibfield  {author} {\bibinfo {author} {\bibfnamefont {D.}~\bibnamefont
  {Haughey}}\ and\ \bibinfo {author} {\bibfnamefont {J.~C.}\ \bibnamefont
  {Earnshaw}},\ }\href@noop {} {\bibfield  {journal} {\bibinfo  {journal}
  {Colloids and Surfaces A: Physicochemical and Engineering Aspects}\ }\textbf
  {\bibinfo {volume} {136}},\ \bibinfo {pages} {217 } (\bibinfo {year}
  {1998})}\BibitemShut {NoStop}%
\bibitem [{\citenamefont {Haynes}(2013)}]{haynes2013crc}%
  \BibitemOpen
  \bibfield  {author} {\bibinfo {author} {\bibfnamefont {W.~M.}\ \bibnamefont
  {Haynes}},\ }\href@noop {} {\emph {\bibinfo {title} {CRC handbook of
  chemistry and physics}}}\ (\bibinfo  {publisher} {CRC press},\ \bibinfo
  {year} {2013})\BibitemShut {NoStop}%
\bibitem [{\citenamefont {Petrache}\ \emph {et~al.}(1998)\citenamefont
  {Petrache}, \citenamefont {Gouliaev}, \citenamefont {Tristram-Nagle},
  \citenamefont {Zhang}, \citenamefont {Suter},\ and\ \citenamefont
  {Nagle}}]{petrache(pre1998)}%
  \BibitemOpen
  \bibfield  {author} {\bibinfo {author} {\bibfnamefont {H.~I.}\ \bibnamefont
  {Petrache}}, \bibinfo {author} {\bibfnamefont {N.}~\bibnamefont {Gouliaev}},
  \bibinfo {author} {\bibfnamefont {S.}~\bibnamefont {Tristram-Nagle}},
  \bibinfo {author} {\bibfnamefont {S.}~\bibnamefont {Zhang}}, \bibinfo
  {author} {\bibfnamefont {R.~M.}\ \bibnamefont {Suter}}, \ and\ \bibinfo
  {author} {\bibfnamefont {J.~F.}\ \bibnamefont {Nagle}},\ }\href@noop {}
  {\bibfield  {journal} {\bibinfo  {journal} {Physical Review E}\ }\textbf
  {\bibinfo {volume} {57}},\ \bibinfo {pages} {7014} (\bibinfo {year}
  {1998})}\BibitemShut {NoStop}%
\bibitem [{\citenamefont {Mallikarjunaiah}\ \emph {et~al.}(2011)\citenamefont
  {Mallikarjunaiah}, \citenamefont {Leftin}, \citenamefont {Kinnun},
  \citenamefont {Justice}, \citenamefont {Rogozea}, \citenamefont {Petrache},\
  and\ \citenamefont {Brown}}]{mallikarjunaiah2011solid}%
  \BibitemOpen
  \bibfield  {author} {\bibinfo {author} {\bibfnamefont {K.}~\bibnamefont
  {Mallikarjunaiah}}, \bibinfo {author} {\bibfnamefont {A.}~\bibnamefont
  {Leftin}}, \bibinfo {author} {\bibfnamefont {J.~J.}\ \bibnamefont {Kinnun}},
  \bibinfo {author} {\bibfnamefont {M.~J.}\ \bibnamefont {Justice}}, \bibinfo
  {author} {\bibfnamefont {A.~L.}\ \bibnamefont {Rogozea}}, \bibinfo {author}
  {\bibfnamefont {H.~I.}\ \bibnamefont {Petrache}}, \ and\ \bibinfo {author}
  {\bibfnamefont {M.~F.}\ \bibnamefont {Brown}},\ }\href@noop {} {\bibfield
  {journal} {\bibinfo  {journal} {Biophysical journal}\ }\textbf {\bibinfo
  {volume} {100}},\ \bibinfo {pages} {98} (\bibinfo {year} {2011})}\BibitemShut
  {NoStop}%
\bibitem [{\citenamefont {Nagle}\ and\ \citenamefont
  {Wiener}(1989)}]{NagleWiener89}%
  \BibitemOpen
  \bibfield  {author} {\bibinfo {author} {\bibfnamefont {J.}~\bibnamefont
  {Nagle}}\ and\ \bibinfo {author} {\bibfnamefont {M.}~\bibnamefont {Wiener}},\
  }\href@noop {} {\bibfield  {journal} {\bibinfo  {journal} {Mol. Cryst. Liq.
  Cryst.}\ }\textbf {\bibinfo {volume} {144}},\ \bibinfo {pages} {235}
  (\bibinfo {year} {1989})}\BibitemShut {NoStop}%
\end{thebibliography}
\end{document}